\documentclass[aps,twocolumn,amsmath,amssymb,showpacs,floatfix,prb,superscriptaddress,
unsortedaddress]{revtex4}
\usepackage{graphicx}

\begin{document}

\title{Optical Integral in the Cuprates and the Question of Sum Rule Violation}
\author{M. R. Norman}
\affiliation{
Materials Science Division, Argonne National Laboratory, Argonne,
Illinois 60439}
\author{A. V. Chubukov}
\affiliation{Department of Physics, University of Wisconsin, Madison, WI
53706}
\author{E. van Heumen}
\affiliation{University of Geneva, 24, Quai E.-Ansermet, Geneva 4, Switzerland}
\author{A. B. Kuzmenko}
\affiliation{University of Geneva, 24, Quai E.-Ansermet, Geneva 4, Switzerland}
\author{D. van der Marel}
\affiliation{University of Geneva, 24, Quai E.-Ansermet, Geneva 4, Switzerland}
\date{\today}
\begin{abstract}
Much attention has been given to a possible violation of the optical sum
rule in the cuprates, and the connection this might have to
 kinetic energy lowering.  The optical integral is composed of a cut-off
 independent term (whose temperature dependence is a measure of the sum rule
 violation), plus a cut-off dependent term that accounts for the extension of
the Drude peak beyond the upper bound of the integral.  We find that the temperature
 dependence of the optical integral in the normal state
of the cuprates can be accounted for solely by the latter term,
implying that the dominant contribution to the observed 
sum rule `violation' in the normal state is due to the finite cut-off.
This cut-off dependent term 
is well modeled by a theory of electrons interacting with a broad spectrum of bosons.
\end{abstract}

\maketitle

The integral of the real part of the optical conductivity with respect to frequency
up to infinity is known as the
f sum rule. It is proportional to $n/m$, and is preserved by charge conservation.\cite{Mahan}  In experiments, however, the conductivity is measured up to a certain frequency cut-off.  In a situation when the system has a single band of low energy carriers, separated by an energy gap from other high energy bands (as in the cuprates), the exact f sum rule reduces to the single band
sum rule of Kubo \cite{Kubo}
\begin{equation}
W = \int_0^{\omega_c} Re \sigma(\omega) d\omega =
f(\omega_c)~\frac{\omega_{pl}^2}{8} \equiv f(\omega_c) ~\frac{\pi e^2 a^2}{2 \hbar^2 V} E_K
\label{1}
\end{equation}
Here
$a$ is the in-plane lattice constant, $V$ the unit cell volume, $\omega_c$ an
ultraviolet cut-off, $\omega_{pl}$ the bare plasma frequency, and
\begin{equation}
E_K=\frac{2}{a^2 N}
\sum_k \frac{\partial^2 \epsilon_k}{\partial k_x^2} n_k
\label{E_k}
\end{equation}
with $N$ the number of k vectors, $\epsilon_k$ the bare dispersion as defined
by the effective single band Hamiltonian, and $n_k$ the momentum
distribution function.  For a Hamiltonian with near neighbor hopping,\cite{HM1}
$E_K$ is equivalent to minus the kinetic energy, $E_{kin} 
\equiv \frac{2}{N} \sum_k \epsilon_k n_k$,
but in general these two quantities differ.\cite{Marsiglio}
 The cut-off  $\omega_c$  is 
assumed to be larger than the bandwidth 
 of the low energy band, but smaller than the gap to other bands.
For this reason, its value in cuprates is typically chosen to be 1 - 1.25 eV.
 $f(\omega_c)$ accounts for the cut-off dependence, which arises from the presence
 of Drude spectral weight beyond $\omega_c$ \cite{Maksimov} and 
 is unity if we formally set $\omega_c$ to infinity while ignoring the interband transitions.
 
Although the f sum rule  is preserved, $W(\omega_c, T)$
in general is not a conserved quantity since both $f(\omega_c)$ \cite{Maksimov}
and $n_k$ \cite{Mike00} depend on $T$.
The $T$ variation of $W$ has been termed the `sum rule violation'.
In conventional superconductors, the sum rule is preserved within experimental 
accuracy.\cite{Tinkham}
In the cuprates, however,  the c-axis conductivity indicates a pronounced violation of the sum 
rule.\cite{Basov}  More recently, similar behavior was found for the in-plane conductivity.\cite{Molegraaf,Santander,Bi2223,Erik}

The reported violation takes two forms.
First, as the temperature is lowered in the normal state, the optical integral increases
 in magnitude.  Then, below $T_c$,
 there is an additional change in the optical integral compared to
 that of the extrapolated normal
 state.  For overdoped compounds, this change is a  decrease,\cite{Deutscher,Carbone}
 but for underdoped and optimal doped compounds, two groups \cite{Molegraaf,Santander,Bi2223,Erik}
found  an increase, though other groups found either no additional effect \cite{Homes,Ortolani}
or a decrease.\cite{Boris}

The finite cut-off was taken into account in several theoretical analyses of the $T$ 
dependence of the optical integral, for instance work
based on the Hubbard model,\cite{Toschi} the t-J model \cite{Carbone} and
the d-density wave model.\cite{Benfatto}
In Ref.~\onlinecite{Maksimov}, the effect of the cut-off 
was considered in the context of electrons coupled to phonons.  The goal of the
present paper is to study the influence of the cut-off on the optical integral
for a model of electrons interacting with a broad spectrum of bosons that two of us
have used previously to model optics data.\cite{MNAC}

In a Drude model, $\sigma(\omega) = (\omega_{pl}^2/4\pi)/(1/\tau-i\omega)$.
From Eq.~\ref{1}, we see that the true sum rule violation is encoded in the $T$
dependence of $\omega_{pl}$.  Although $\omega_{pl}$ is well known
to be a strong function of doping,\cite{RMP} the question of its $T$
dependence is more subtle because of the presence of $f(\omega_c)$. 
Integrating over $\omega$ and expanding for  $\omega_c \tau >> 1$, we obtain
$W(\omega_c) = \frac{\omega_{pl}^2}{8} f(\omega_c)$, where
$f(\omega_c) = (1-\frac{2}{\pi}\frac{1}{\omega_c \tau})$.
For infinite cut-off, $f(\omega_c) =1$ and  $W = \omega_{pl}^2/8$, 
but for a finite cut-off, $f(\omega_c)$ is  
 a constant minus a term proportional
to $1/\omega_c \tau$.  If $1/\tau$ changes with $T$, then one obtains 
a sum rule `violation' even if $\omega_{pl}$ is $T$ independent.\cite{Maksimov}

In general, optical integral changes due to  $E_K$ (i.e., $\omega_{pl}$)
 and $f(\omega_c)$ are both present, and 
 the difference of optical integrals at two different temperatures, 
$\Delta W = W(T_1) - W(T_2)$, goes as
\begin{equation}
\Delta W = \alpha \Delta E_K + \beta \Delta f(\omega_c)
\label{4}
\end{equation}
where $\alpha$ and $\beta$ are constants. 
The issue then is which term contributes more to the sum rule violation at a given $\omega_c$. If the variation predominantly comes from $E_K$, it 
  would be a true sum rule violation, related  to the variation of the kinetic energy.  The increase of $W$ with decreasing $T$ would then imply that the kinetic energy decreases with decreasing $T$.
 If the change of $W$ comes from $f(\omega_c)$, the sum-rule `violation'
 would be a cut-off effect, unrelated to the behavior of the kinetic energy. 

 In this paper, for simplicity, we concentrate on the temperature variation of the optical 
 integral in the normal state.  We find that the data can be fit by Eq.~\ref{4} with $\Delta E_K=0$.
Based on the accuracy of the fit,  we estimate that the true sum rule violation, $\Delta E_K$, must be smaller than  $\sim20\%$ of $\Delta W$.
Moreover, we find that the temperature variation due to the second term in Eq.~\ref{4},
and its dependence on the cut-off,
is well modeled by a theory of fermions  
 interacting with a broad spectrum of bosons. 

We considered two models for the bosonic spectrum.
The first is a `gapped' marginal Fermi liquid, 
where the spectrum is flat in frequency up to an 
upper cut-off $\omega_2$ \cite{MFL}
\begin{equation}
\alpha^2F(\Omega)_{GMFL} = \frac{\Gamma}{\omega_2-\omega_1}
\Theta(\Omega-\omega_1)\Theta(\omega_2-\Omega)
\end{equation}
with a lower cut-off  $\omega_1$ put in by hand to prevent divergences.
The second is a Lorentzian spectrum typical for overdamped
 spin and charge fluctuations \cite{MMP}
\begin{equation}
\alpha^2F(\Omega)_{Lor} = \frac{\Gamma \Omega}{\gamma^2+\Omega^2}
\label{17}
\end{equation}

The computational procedure is straightforward: 
$\alpha^2F$ is used to calculate the single-particle self-energy, and from
this the current-current response function to obtain the conductivity.
The computational  procedure can be simplified, as shown by Allen,\cite{Allen} by presenting $\sigma (\omega)$ in a generalized Drude form 
\begin{equation}
\sigma(\omega) = \frac{\omega_{pl}^2}{4\pi}\frac{1}{1/\tau(\omega)-i\omega m^*(\omega)}
\end{equation}
and  approximating $1/\tau (\omega)$ by \cite{Shulga}
\begin{eqnarray}
1/\tau(\omega) = 2\Gamma_i + \frac{1}{\omega} \int^{\infty}_0 d\Omega \alpha^2F(\Omega)
[2\omega \coth(\frac{\Omega}{2T}) \nonumber \\
- (\omega +\Omega) \coth(\frac{\omega+\Omega}{2T})
+ (\omega -\Omega) \coth(\frac{\omega-\Omega}{2T})]
\label{tau}
\end{eqnarray}
where $2\Gamma_i$ is the impurity contribution.\cite{foot2}
 For electrons interacting with a broad spectrum of
bosons, this approximation is essentially identical to the exact Kubo 
result.\cite{MNAC} The optical mass, $m^* (\omega)$, can then 
be determined by a Kramers-Kronig transformation of $1/\tau (\omega)$.  

One can show quite generally that for an arbitrary form of $\alpha^2 F (\Omega)$, 
$W(\omega_c, T)$ 
asymptotically approaches $W(\infty)$ 
as $W(\omega_c, T) = W(\infty) (1 - (8/\pi) A(T)/\omega_c)$, where
$A(T) = \int_0^\infty d \Omega \alpha^2 F(\Omega) n_B (\Omega)$ with $n_B$ the Bose function.
At high $T$,  therefore, $W(\omega_c, T) - W(\infty)$  scales as $T$ for arbitrary $\omega_c$.
This asymptotic behavior, however, sets in 
for $\omega_c$ much larger than the upper cut-off in $\alpha^2 F (\Omega)$.
This behavior would adequately describe the data if the bosonic spectrum sits at low frequencies
 as for phonons,\cite{Maksimov} but this does not appear to be the case in the cuprates,
 where the inferred bosonic spectrum from the infrared data extends to quite high 
 frequencies.\cite{Hwang,EVH,MNAC}
For  comparison with
 experiment, therefore, we need to know $W(\omega_c, T)$ not only for arbitrary $T$,
but also for $\omega_c$ which are only a few times larger than the energy range of $\alpha^2 F$.

\begin{figure}
\includegraphics[width=\columnwidth] {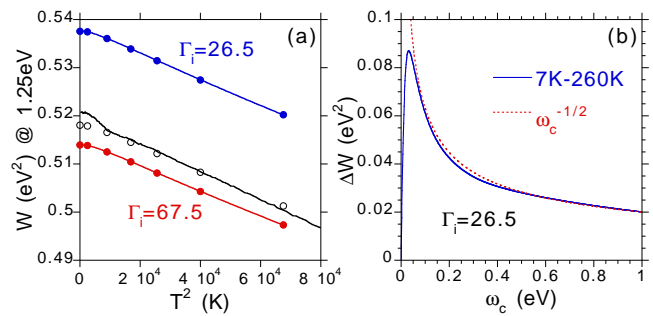}
\caption{(Color online) (a) Optical integral, $W$, for a 1.25 eV cut-off versus $T^2$ for the gapped
marginal Fermi liquid model ($\Gamma$=270.5 meV, $\omega_1$=15.5 meV, $\omega_2$=301 meV)
for two different values of the impurity scattering, $\Gamma_i$, with $\omega_{pl}$=2.4 eV.
The curves are fits to the calculated solid dots using the $T$ dependence of Eq.~\ref{16}.
The middle curve is the
experimental $W$ of Ref.~\onlinecite{Molegraaf} for a Bi$_2$Sr$_2$CaCu$_2$O$_8$
sample with a $T_c$ of 88K.  The open
circles are the $\Gamma_i$=26.5 meV case with $\omega_{pl}$=2.356 eV instead.
(b) Difference of the calculated optical integrals ($T$=7K minus $T$=260K) versus the cut-off
for $\Gamma_i$=26.5 meV.
The dashed line is a $1/\sqrt{\omega_c}$ fit (with $\Delta E_K$=0).}
\label{fig1}
\end{figure}

\begin{figure}
\includegraphics[width=\columnwidth] {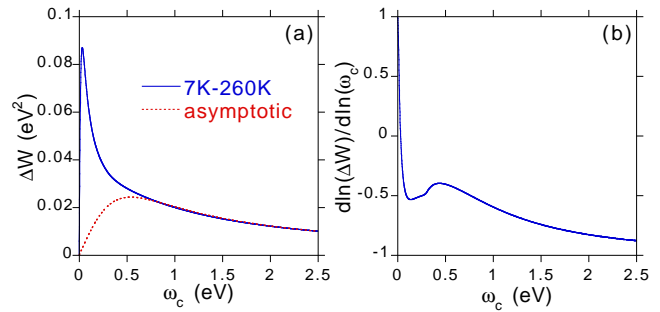}
\caption{(Color online) (a) Optical integral difference from Fig.~1b versus the cut-off as compared
to the asymptotic expression of Eq.~\ref{16}. 
(b) Logarithmic derivative of $\Delta W$ versus the cut-off.}
\label{fig2}
\end{figure}

 We start with the `gapped' marginal Fermi liquid model.  The parameters $\Gamma$, $\omega_1$ and $\omega_2$ 
were chosen \cite{MNAC} so as to fit the data of Ref.~\onlinecite{vdM} at one particular temperature.
We do not optimize them for the data we compare to here, in order to demonstrate the generality
of our arguments.
$\omega_2$ is essentially equal to the peak frequency in the real part of the optical
self-energy, $(m^*(\omega)-1)\omega$, whereas $\Gamma$ is set by the overall size of the optical
self-energy.
We treat $\alpha^2F$ and $\omega_{pl}$ as $T$ independent, so as to concentrate exclusively
on the effect
of $\omega_c$, though the actual quantities may depend on $T$.\cite{Hwang,EVH}
As a consequence, the only thermal
effects which enter are the $\coth$ factors 
 in $1/\tau (\omega)$ in Eq.~\ref{tau}.
 In Fig.~1a, we show the variation of the calculated optical integral with $T$ for two different
values of $\Gamma_i$, and compare this to the data of Ref.~\onlinecite{Molegraaf}.
The results above $T_c$ are consistent with a behavior that goes as a constant minus
 a $T^2$ term in both the data and the theory.  Moreover, the $T^2$ slopes are identical (the
 relative shift in $W$ can be matched by small changes in either $\Gamma_i$
 or $\omega_{pl}$).
In Fig.~1b, we show the difference of the calculated optical integrals at two different $T$
versus $\omega_c$.  After an initial rise (due to the fact that the Drude peak is narrower at 
lower $T$),
 the difference decays.  Unlike the simple Drude model where this decay
goes as $1/\omega_c$, the decay appears to be more like $1/\sqrt{\omega_c}$
 for cut-offs ranging from 0.1 eV to 1 eV.
 To obtain more insight, we show in Fig.~2b the logarithmic derivative 
of $\Delta W$ versus $\omega_c$.  The approximate
-1/2 power is an intermediate frequency result, and one can see the approach
to the asymptotic power of -1  for very large cut-offs.

The behavior for large $\omega_c$ can be also studied analytically.
To start, we  rewrite the optical integral as
\begin{equation}
W(\omega_c, T) = W(\infty) - 
\int_{\omega_c}^\infty Re \sigma (\omega, T)  d \omega
\end{equation}
where $W(\infty) = \omega^2_{pl}/8$.  We then 
note \cite{MNAC} that for $\omega$ larger than the upper cut-off $\omega_2$ of the gapped marginal 
Fermi liquid model, 
\begin{equation}
1/\tau = 1/\tau_{high} = 2\Gamma_i + \frac{\Gamma}{\omega_2-\omega_1}
(4T\ln\frac{\sinh\frac{\omega_2}{2T}}{\sinh\frac{\omega_1}{2T}} - \frac{\omega_2^2-\omega_1^2}{\omega})
\end{equation}
For $T << \omega_2$ and $\omega_1 << \omega_2$ (which are always satisfied),
 this reduces to
\begin{equation}
1/\tau_{high} = 1/\tau_0 - \frac{4\Gamma T}{\omega_2}
\ln(1-e^{-\omega_1/T})
\end{equation}
where $1/\tau_0 = 2\Gamma_i + 2\Gamma - \Gamma \omega_2/\omega$.
Ignoring the frequency dependence of $1/\tau_0$, and setting $m^*$ to 1
($\omega_c >> \omega_2$),\cite{foot3} we then obtain
$\Delta W(\omega_c) = 
\frac{\omega_{pl}^2}{4\pi} \Delta (\tan^{-1}(\omega_c\tau_{high}))$
where again $\Delta W(\omega_c)=W(\omega_c,T_1)-W(\omega_c,T_2)$.
Expanding in $\Delta T$, we obtain
\begin{equation}
\Delta W(\omega_c) = \frac{\omega_{pl}^2}{2\pi\omega_2} \frac{\omega_c^*}{1+(\omega_c^*)^2}
\Delta \left(T \ln(1-e^{-\omega_1/T})\right)
\label{16}
\end{equation}
where $\omega_c^*=\omega_c/(2\Gamma)$.
In Fig.~2a, we plot Eq.~\ref{16} versus the calculated optical integral difference, and see that they
match for cut-offs beyond 0.7 eV.  Moreover, the $T$ dependence of
 Eq.~\ref{16}
matches the $T$ evolution of the optical integral,
 as can be seen in Fig.~1a,
and so a $T^2$ behavior is only approximate.
The true dependence is
 $T \ln(1-e^{-\omega_1/T})$ as in Eq.~\ref{16}, however this is very close to $T^2$ over a wide range of temperatures.    

The above analysis can also be performed for the Lorentzian model
(the numerical results are similar to Fig.~1, and we do not present them here).
Extending  the analysis in Ref.~\onlinecite{MNAC} 
 to finite temperatures, we obtain
\begin{equation}
1/\tau (T) = 1/\tau_0 + 4 \Gamma \int_0^\infty \frac{x dx}{x^2 +1}~\frac{1}{e^{x/T^*}-1}
\end{equation}
where
$1/\tau_0 = 2 \Gamma \ln \frac{\sqrt{\Omega^2_c + \gamma^2}}{\gamma}$
with $T^* = T/\gamma$, and $\Omega_c$ is an upper cut-off for $\alpha^2 F_{Lor}$.
 Assuming that $\Omega_c >> \gamma$, we have
$\Delta W(\omega_c) = \frac{\omega_{pl}^2}{4\pi} \Delta (\tan^{-1}(\omega_c\tau))$.
Expanding around $T=0$, we obtain
\begin{equation}
W(\omega_c, T) \approx W(\omega_c,T=0) -  \frac{\omega_{pl}^2}{4\pi}~C~ (T^*)^2
\end{equation}
where
\begin{equation}
C = \frac{\pi^2}{6}~\frac{\omega_c}{\Gamma \ln^2 (\Omega_c/\gamma)} ~\frac{1}{1 + \left(\frac{\omega_c}{2\Gamma \ln (\Omega_c/\gamma)}\right)^2}
\end{equation}
This time, we find a truly quadratic behavior in $T$,\cite{foot4} which is a consequence of the
fact that $\alpha^2F_{Lor}$ is linear in $\omega$ at small $\omega$. 
 The dependence of $\Delta W$ 
 on the frequency cut-off is the same as the gapped marginal Fermi liquid
model, except that the quantity $\omega_c^*$ in 
 Eq.~\ref{16} is now $\omega_c/(2\Gamma\ln(\Omega_c/\gamma))$.
 
\begin{figure}
\includegraphics[width=\columnwidth] {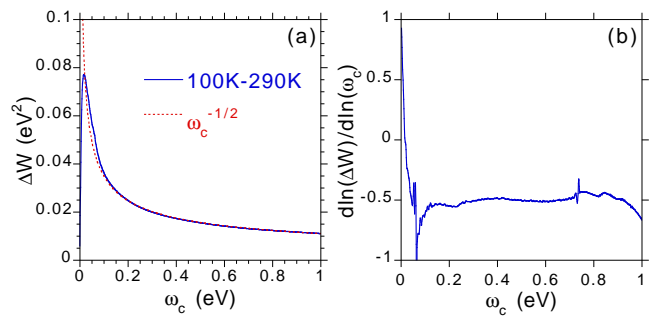}
\caption{(Color online) (a) Optical integral difference for a HgBa$_2$CuO$_4$ sample
with a $T_c$ of 97K (data from Ref.~\onlinecite{Erik}).  The dashed line is a $1/\sqrt{\omega_c}$ fit
(with $\Delta E_K$=0).
(b) logarithmic derivative of $\Delta W$ versus $\omega_c$.}
\label{fig3}
\end{figure}

We now return to experiment.
In Fig.~1a, we plot the experimental optical integral for a 1.25 eV cut-off
from the data of Ref.~\onlinecite{Molegraaf} versus our calculations.  The magnitude and $T$
variation of $W$ are essentially equivalent to these calculations, which assumed a $T$
independent $\omega_{pl}$.\cite{foot1}  In Fig.~3a, we show the
 difference between the measured optical integrals 
for two temperatures versus the frequency cut-off from the data of Ref.~\onlinecite{Erik}.  A 
$1/\sqrt{\omega_c}$ dependence, with a {\it zero} offset, fits the data quite well,
as with the theory in Fig.~1b.
This is further demonstrated by the logarithmic derivative, as plotted in Fig.~3b.
From these observations,  we conclude that  the dominant contribution to the $T$ dependence of the optical integral in the
normal state can be  attributed to the finite cut-off. 
The true sum rule `violation' term $\Delta E_K$  is estimated to be no larger than $\sim20\%$ of $\Delta W$, as noted above.
Although we do not expect our analysis to be the entire story, in that
 there is experimental evidence that $\alpha^2F$ is
$T$ dependent,\cite{Hwang} even though this dependence is weak in the normal state,\cite{EVH}
still, based on the good agreement of the calculations with
experiment, we would argue that  the
bulk of the observed $T$ dependence in the normal state is related to the finite cut-off.

The above analysis is non-trivial to extend to below $T_c$, as this requires some assumptions about the pairing kernel,  since one needs to construct 
 the anomalous Green's function, $F$, in order to evaluate the current-current response function. 
The additional increase of $W(\omega_c, T)$ below $T_c$ in optimal and
underdoped cuprates, reported in Refs.~\onlinecite{Molegraaf,Santander,Bi2223,Erik}, could be due to
the strong decrease in $1/\tau$ observed by a variety of probes.
On the other hand, strong coupling calculations cast doubt on
 a cut-off explanation, as the influence of $f(\omega_c)$
would be to give rise to a negative $W_{sc} - W_n$  for cut-offs near 1 eV.\cite{sign}
Moreover, similar strong coupling calculations
 of the variation of $E_K$ between the normal and superconducting states yield a positive $E^{sc}_K$ - $E^n_K$ $\sim$ 1 meV for the underdoped case,\cite{NP}
which  is consistent both in sign and magnitude 
 with the results of Refs.~\onlinecite{Molegraaf,Santander,Bi2223,Erik}. 
This  implies that there may be a true sum rule violation below $T_c$.

The authors would like to thank Nicole Bontemps for suggesting this study, and Frank Marsiglio for useful conversations.
MN was supported by the U.~S.~Dept.~of Energy, Office of Science,
under Contract No.~DE-AC02-06CH11357, and
AC by NSF-DMR  0604406.
The work at the University of Geneva is supported by the Swiss NSF
through grant 200020-113293 and the National Center of
Competence in Research (NCCR) Materials with Novel Electronic
Properties-MaNEP.
 MN and AC thank the
Aspen Center for Physics where this work was completed.


\begin{thebibliography}{99}
\bibitem{foot2}
Our definition of $\alpha^2F$ is that of Ref.~\onlinecite{MNAC}, and differs by
a factor of $\pi$ from most studies.
\bibitem{foot1}
If we allowed for a realistic (i.e., non-free electron) band dispersion, then
$\Delta W$ for these two models would also yield a non-zero 
$\Delta E_K$, as in Ref.~\onlinecite{NP}.
\end{thebibliography}
\end{document}